%% file: EXO-14-006_temp.tex
\pdfoutput=1

\documentclass[11pt,twoside,a4paper,cmspaper,final,collab]{cms-tdr}
\def\svnVersion{426646P}\def\cmsCernNoTag{CERN-EP-2016-295}\def\cmsCernDate{\today}\def\cmsMessage{Published in Physics Letters B as \href{http://dx.doi.org/10.1016/j.physletb.2017.08.069}{\doi{10.1016/j.physletb.2017.08.069.}}}

\begin{document}\cmsNoteHeader{EXO-14-006}

\hyphenation{had-ron-i-za-tion}
\hyphenation{cal-or-i-me-ter}
\hyphenation{de-vices}
\RCS$Revision: 426555 $
\RCS$HeadURL: svn+ssh://svn.cern.ch/reps/tdr2/papers/EXO-14-006/trunk/EXO-14-006.tex $
\RCS$Id: EXO-14-006.tex 426555 2017-09-22 05:48:44Z hdyoo $
\newlength\cmsFigWidth
\ifthenelse{\boolean{cms@external}}{\setlength\cmsFigWidth{\columnwidth}}{\setlength\cmsFigWidth{0.85\textwidth}}
\ifthenelse{\boolean{cms@external}}{\providecommand{\cmsLeft}{top}}{\providecommand{\cmsLeft}{left}}
\ifthenelse{\boolean{cms@external}}{\providecommand{\cmsRight}{bottom}}{\providecommand{\cmsRight}{right}}

\newcommand{\mumumumu}{\ensuremath{\Pgm\Pgm\Pgm\Pgm}\xspace}
\newcommand{\mumumue}{\ensuremath{\Pgm\Pgm\Pgm \Pe}\xspace}
\newcommand{\mumuee}{\ensuremath{\Pgm\Pgm \Pe\Pe}\xspace}
\newcommand{\mueee}{\ensuremath{\Pgm\Pe\Pe\Pe}\xspace}
\newcommand{\eeee}{\ensuremath{\Pe\Pe\Pe\Pe}\xspace}
\newcommand{\mZp}{\ensuremath{m_{\PZpr}}\xspace}
\newcommand{\mphi}{\ensuremath{m_{\varphi}}\xspace}
\newcommand{\mfl}{\ensuremath{m_{4\ell}}\xspace}

\cmsNoteHeader{EXO-14-006}
\title{Search for leptophobic \PZpr bosons decaying into four-lepton final states
in proton-proton collisions at $\sqrt{s} = 8\TeV$}

\date{\today}

\abstract{
A search for heavy narrow resonances decaying into four-lepton final states
has been performed using proton-proton collision data at $\sqrt{s} = 8\TeV$
collected by the CMS experiment, corresponding to an integrated luminosity of 19.7\fbinv.
No excess of events over the standard model background expectation is observed.
Upper limits for a benchmark model
on the product of cross section and branching fraction for the production
of these heavy narrow resonances are presented.
The limit excludes leptophobic \PZpr bosons with masses below 2.5\TeV within the benchmark model.
This is the first result to constrain a leptophobic \PZpr resonance in the four-lepton channel.
}

\hypersetup{%
pdfauthor={CMS Collaboration},%
pdftitle={Search for leptophobic Z' bosons decaying into four-lepton final states in proton-proton collisions at sqrt(s) = 8 TeV},%
pdfsubject={CMS},%
pdfkeywords={LHC, CMS, physics, exotica, Z', four leptons}}

\maketitle

\section{Introduction}

Extensions of the standard model (SM) that incorporate one or more extra Abelian gauge groups
predict the existence of one or more neutral gauge bosons~\cite{GUT1, GUT2}.
These occur naturally in most grand unified theories.
Heavy neutral bosons are also predicted in models with extra spatial dimensions~\cite{KK1, KK2},
\eg Randall--Sundrum models~\cite{Ext1, Ext2}, where these resonances may arise from
Kaluza--Klein excitations of a graviton.
Searches for heavy neutral resonances at hadron colliders, and most recently at the CERN LHC, are typically performed using the
dijet~\cite{ATLASdijet, ATLASdijet13, CMSdijet, CMSdijet13}, dilepton~\cite{ATLASZp1, ATLASZp2, EXO-12-061, EXO-15-005},
diphoton~\cite{ATLASdiphoton1, CMSdiphoton1, CMSdiphoton2},
diboson~\cite{ATLASdib1, ATLASdib2, ATLASdib3, EXO-16-021, EXO-16-025, EXO-12-053, EXO-14-009},
and \ttbar~\cite{ATLASttbar1, ATLASttbar2, CMSttbar1, CMSttbar2} final states.
The dilepton channel provides a clean signal compared with the dijet and \ttbar channels.
However, in leptophobic \PZpr~models, where the \PZpr does not couple to SM leptons, the dilepton limits are not applicable.
Although searches based on the dijet final state remain applicable, they suffer from large
dijet background produced by quantum chromodynamics (QCD) subprocesses.
We extend the search for heavy neutral vector bosons by considering possible \PZpr decays into
new particles predicted by various theoretical extensions of the SM.

\begin{figure}[htbp]
\centering
\includegraphics[width=0.5\textwidth]{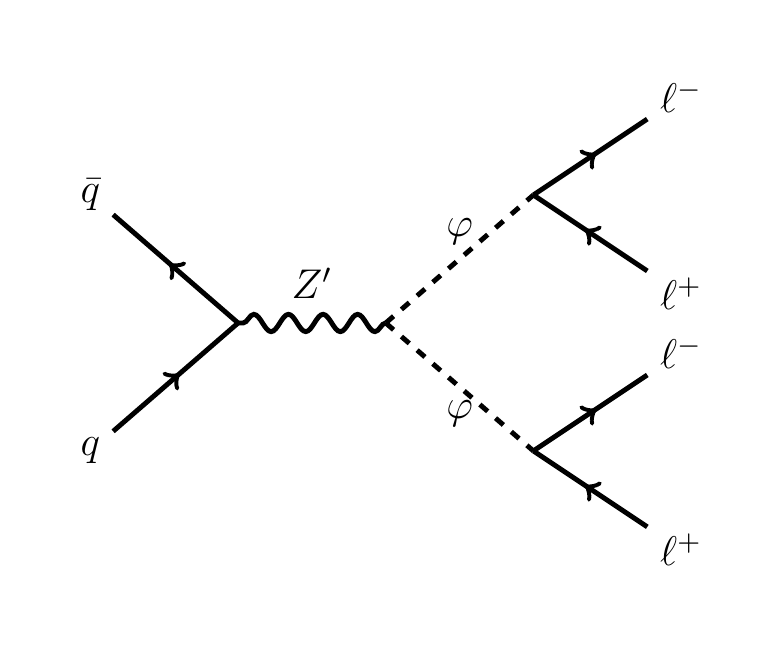}
\caption{
    Leading order Feynman diagram for the production and cascade decay of a \PZpr resonance to a four-lepton final state.
}
\label{fig:decayMode}

\end{figure}

In this Letter, we report on a search for a leptophobic \PZpr resonance
that decays into four leptons via cascade decays as described in Ref.~\cite{HSLee:FourLepton}.
In this model, the \PZpr is coupled to quark pairs but not to lepton pairs,
and can be produced with a large cross section at the LHC.
These non-standard \PZpr resonances also decay to pairs of new scalar bosons ($\varphi$) each of which subsequently decays to
pairs of leptons ($\varphi \to \ell\ell'$, where $\ell\text{ and }\ell^\prime = \Pe$ or $\Pgm$).
Figure~\ref{fig:decayMode} shows the leading-order Feynman diagram for the production
of four-lepton final states via a \PZpr resonance at a hadron collider.
The reconstruction of the $\varphi$ bosons in the dilepton channel is inefficient
if the difference between \PZpr and $\varphi$ masses is large
and the two daughter leptons are consequently highly collimated.
In the following sections we describe a technique to increase the selection efficiency.

The analysis is a search for heavy narrow resonances decaying into four isolated final state leptons.
The benchmark model~\cite{HSLee:FourLepton} assumes ($\Gamma/M<1\%$), corresponding to a natural width
of the \PZpr resonance that is much smaller than the detector resolution.
The following final states are considered: $\mumumumu$, $\mumumue$, $\mumuee$, $\mueee$, and $\eeee$.
The $\mumuee$, $\mumumue$ and $\mueee$ channels are included to allow for the possibility of lepton
flavor violation (LFV)~\cite{LFV1,LFV2,LFV3} in the decays of the new scalar bosons.
In this Letter, we set limits on the product of the cross section and branching fraction for production
and decay to four leptons, and interpret the results in the context of the benchmark model described above~\cite{HSLee:FourLepton}.

\section{The CMS detector}

The central feature of the CMS apparatus is a superconducting solenoid of 6\unit{m} internal diameter,
providing a magnetic field of 3.8\unit{T}. Within the solenoid volume are a silicon pixel
and strip tracker, a lead tungstate crystal electromagnetic calorimeter (ECAL), and a brass and scintillator
hadron calorimeter (HCAL). Each detector is composed of a barrel and two endcap sections.
Muons are measured in gas-ionization detectors embedded in the steel flux-return yoke outside the solenoid.
Extensive forward calorimetry complements the coverage provided by the barrel and endcap detectors.

Muons are measured in the range $\abs{\eta}< 2.4$ with detection planes made using
three technologies: drift tubes, cathode strip chambers, and resistive-plate chambers.
Matching muons to tracks measured in the silicon tracker results in a relative
$\pt$ resolution for muons with $20<\pt<100\GeV$ of 1.3--2.0\% in the
barrel and better than 6\% in the endcaps.
The \pt resolution in the barrel is better than 10\% for muons with \pt up to 1\TeV~\cite{CMS-PAPER-MUO-10-004}.

The ECAL consists of 75\,848 crystals that
provide coverage in pseudorapidity $\abs{ \eta }< 1.48 $ in a barrel region (EB)
and $1.48 <\abs{ \eta } < 3.00$ in two endcap regions (EE).
The electron momentum is estimated by combining the energy measurement in the ECAL with the
momentum measurement in the tracker. The momentum resolution for electrons with transverse momentum $\pt\approx45\GeV$
from $\Z \to \Pep \Pem$ decays ranges from 1.7\% for nonshowering electrons (approximately 30\%)
in the barrel region
to 4.5\% for showering electrons (approximately 60\%) in the endcaps~\cite{CMS:EGM-13-001}.

A more detailed description of the CMS detector, together with a definition of the coordinate system
used and the relevant kinematic variables, can be found in Ref.~\cite{Chatrchyan:2008zzk}.

\section{The simulated event samples}

The Monte Carlo (MC) generator program used to produce the simulated event samples for
the benchmark model is \CALCHEP
3.4.1~\cite{CalcHep} interfaced with \PYTHIA 6.4.24~\cite{Sjostrand:2006za}.
These samples are divided into five decay channels ($\mumumumu, \mumumue, \mumuee$, $\mueee, \eeee$)
for different \PZpr boson masses (\mZp) ranging from 250 to 3000\GeV in increments of 250\GeV.
The benchmark model assumes that new particles other than \PZpr and $\varphi$ are
heavy enough not to affect the production and decay of the \PZpr boson.
Signal MC samples are produced with six different values of the $\varphi$ mass (\mphi),
with \mphi = 50\GeV used as the reference mass value in the interpretation of the results.
An important feature of this analysis is the presence of a ``boosted signature'' associated with
the collimation of the two leptons coming from the same parent particle and
resulting from the large difference between \mZp and \mphi.
In addition, samples are generated with \mphi masses of 5, 10, 20, 30 and 40\% of \mZp,
for which, in most cases, the contribution from the boosted signature is less important.
The product of the leading order (LO) signal cross section and branching fraction
in each channel varies with \mZp (from 250 to 3000\GeV) as follows:
$\mumumumu$ and $\eeee$ from $0.8$\unit{pb} to $3.0\times10^{-6}$\unit{pb}, $\mumuee$ from $12.3$\unit{pb} to $4.7\times10^{-5}$\unit{pb},
and $\mumumue$ and $\mueee$ from $3.1$\unit{pb} to $1.2\times10^{-5}$\unit{pb}.
The branching fraction of $\varphi \to \ell\ell^\prime$ is set to 1 and therefore only the leptonic decay channels are considered.
These signal MC samples are used to optimize event selection, evaluate signal efficiencies and calculate exclusion limits.

The dominant SM background is the production of $\cPZ\cPZ$ decaying into four leptons.
The $\qqbar$-induced $\cPZ\cPZ$ production is generated using the \PYTHIA event generator
and the $\cPg\cPg$-induced production using the \textsc{gg2zz} program~\cite{GG2ZZ}.
Additional backgrounds from diboson production ($\PW\PW$ and $\PW\cPZ$) are generated with \PYTHIA,
and from top quark production (\ttbar, $\PQt\PW$, and $\overline{\cmsSymbolFace{t}}$\PW) are generated with \POWHEG 1.0~\cite{POWHEG}.
Other processes, such as $\ttbar\Z$ and
triboson production ($\PW\PW\gamma$, $\PW\PW\cPZ$, $\PW\cPZ\cPZ$, and $\cPZ\cPZ\cPZ$),
are generated with \MADGRAPH 5.1.3.30~\cite{Madgraph}.
Simulated event samples are normalized using the integrated luminosity
and higher order theoretical cross sections: next-to-next-to-leading order for \ttbar~\cite{NNLOttbar}
and next-to-leading order for $\cPZ\cPZ$~\cite{NLOZZ} and the other backgrounds.

The MC samples are generated using the CTEQ6L~\cite{CTEQ6L} set of parton distribution functions (PDFs)
and the \PYTHIA Z2* tune~\cite{Z2stune1, Z2stune2} in order to model the proton structure and the underlying event.
The samples are then processed with the full CMS detector simulation software, based on \GEANTfour~\cite{Geant4, Geant4da},
which includes trigger simulation and event reconstruction.

\section{Event selection}

The 2012 data set of proton-proton collisions at $\sqrt{s} = 8\TeV$, corresponding to an integrated
luminosity of $19.7$\fbinv, is used for the analysis.
Data are collected with lepton triggers with various \pt thresholds.
The trigger used for the muon-enriched channels ($\mumumumu, \mumumue$) requires the presence of at least one muon
candidate with $\pt>40\GeV$ and $\abs{\eta}<2.1$.
The trigger used for the electron-enriched channels ($\mueee, \eeee$) requires two clusters of energy deposits in the ECAL
with transverse energy $\et >33\GeV$ each.
For the $\mumuee$ channel, the trigger requires both an electron and a muon with $\pt>22\GeV$.

In the subsequent analysis, events are required to contain a reconstructed primary vertex (PV) with at least four
associated tracks, and its $r$ ($z$) coordinates are required to be within 2\,(24)\unit{cm} of the nominal
interaction point.
The PV is defined as the vertex with the highest sum of $\pt^{2}$ for the associated tracks.
We select the events with four leptons in the final state, where the leptons are  identified by the selection criteria
described below.
The two leading leptons are required to have $\pt>45\GeV$ to ensure that the trigger is fully efficient
for the selected events. This requirement has a negligible effect on the signal acceptance.
The two subleading leptons are required to have $\pt>30\GeV$.
This choice balances loss of efficiency against increased misidentification probability.
All four leptons must satisfy $\abs{\eta}<2.4$. No charge requirement is applied to the lepton selection.

Muon candidates are reconstructed by a combined fit including hits in both tracking and muon detectors
(``global muons'')~\cite{CMS-PAPER-MUO-10-004}.
The tracks associated with global muons are required to have the following properties:
at least one pixel detector hit, at least six strip tracker layers with hits,
at least one muon chamber hit, at least two muon detector planes with muon segments, a transverse impact parameter of
the tracker track $\abs{d_{xy}}<0.2$\unit{cm} with respect to the PV, a longitudinal distance of the tracker track
$\abs{d_z}<0.5$\unit{cm} with respect to the PV, and $\delta\pt/\pt<0.3$ where $\delta\pt$ is the uncertainty
in the measured \pt of the track.
All muon candidates are required to be isolated.
A muon is considered isolated if the scalar
\pt sum of all tracks within a cone of $\Delta R<0.3$ around the muon,
excluding the muon candidate itself, does
not exceed 10\% of the muon \pt, where $\Delta R = \sqrt{\smash[b]{(\Delta\phi)^2 + (\Delta\eta)^2}}$.
If there is a second lepton candidate within a cone $\Delta R<0.3$, we remove its contribution.

An electron candidate is identified by matching a cluster in the ECAL to a track in the silicon tracker~\cite{CMS:EGM-13-001}.
Identification criteria are applied to suppress jets misidentified as electrons.
Electrons are required to pass the
following criteria: the profile of energy deposition in the ECAL should be consistent with an electron,
the sum of HCAL energy deposits behind the ECAL cluster should be less than 10\% of the associated ECAL deposit,
the track associated with the cluster should have no more than one hit missing in the pixel detector layers
and $\abs{d_{xy}}$ should be less than 0.02\unit{cm} with respect to the selected PV.
All electron candidates are required to be isolated using the following definition:
within a cone $\Delta R<0.3$ around the track of the electron candidate, the \pt sum of all other tracks
is required to be less than 5\GeV and the \et sum of the energies of the calorimeter deposits that
are not associated with the candidate is required to be less than 5\% of the candidate's \et.
This differs from the isolation requirement of 3\% in Ref.~\cite{EXO-12-061},
because of the inefficiency (of approximately 6\% at electron \et = 1\TeV)
caused by overlapping electrons due to the high Lorentz boost of the $\varphi$ boson ($\mphi = 50\GeV$).
In addition, if the direction of the second lepton candidate falls within the isolation cone of the first ($\Delta R<0.3$),
the contributions it makes to both \pt and \et are subtracted when imposing the isolation requirements.

The kinematic distributions of the final-state particles are similar for all five channels.
The final state consists of two leading leptons with high \pt and two subleading leptons
with relatively low \pt.
The two leptons from the same parent $\varphi$ boson can be highly Lorentz boosted
if \mphi is significantly smaller than \mZp.
This feature is generally found for high-mass ($\mZp>1\TeV$) samples in the case of $\mphi = 50\GeV$.
This boosted signature introduces a significant inefficiency for the event selection
except for the LFV case ($\varphi$ decaying into $\Pe\Pgm$).
To take into account the boosted signature for $\varphi$ decaying into $\Pgm\Pgm$,
one of the muon candidates selected by the above criteria is allowed to be reconstructed only as a tracker muon,
a track in the tracker matched to track segments in the muon system (``tracker muons'')~\cite{CMS-PAPER-MUO-10-004},
if the two muons are as close as $\Delta R<0.4$.
In such exceptional cases, the requirements of at least one muon chamber hit
and at least two muon detector planes with muon segments
are not applied to the tracker muon.

The boosted signature for a $\varphi$ decaying into $\Pe\Pe$ is much more complicated
since the electrons can easily merge into a single cluster in the ECAL.
In this case, only one electron candidate is reconstructed from the two original electrons.
The probability for having a merged candidate is about 50\% with $\mZp = 3\TeV$ and $\mphi = 50\GeV$.
These events would be rejected by the four-lepton requirement, introducing a large signal inefficiency.
To select such events, an electron candidate having
a ratio of ECAL cluster energy to track momentum larger than 1.5 and a second track with $\pt>30\GeV$ within
a cone of $\Delta R(\text{electron, track})<0.25$, is considered as a ``merged electron''.
Events are accepted with three (two) leptons if they contain one (two) merged electron(s),
since each merged electron is considered to contribute two electrons to the total.
In order to avoid significant misidentification, merged electrons are only considered if the ECAL cluster energy is
bigger than 500\GeV.

The dominant background in this analysis arises from $\cPZ\cPZ$ events decaying into four leptons.
To suppress this background, events with two oppositely charged same-flavor lepton pairs are rejected
if the mass of the lepton pair, $m_{\ell\ell}$, is in the range 89--93\GeV.
The $\cPZ$ mass window is made as narrow as possible in order to minimise degradation of the signal efficiency
in the case of $\mphi\approx m_{\cPZ}$.
This requirement results in negligible signal efficiency loss for $\mZp>500\GeV$.
At lower masses, the efficiency loss increases and is approximately 20\,(7)\% at $\mZp = 250\GeV$
for the $\eeee~(\mumumumu)$ channel.
More than 70\% (30\%) of the $\cPZ\cPZ$ background is rejected by the mass window veto requirement in the muon (electron) channel.
This requirement is not applied to the merged electron case, thus accounting for the difference in rejection efficiency
for the two channels.

The event selection efficiency is 50--70\% ($\mumumumu$), 55--65\% ($\mumumue$ and $\mumuee$) and
45--65\% ($\mueee$ and $\eeee$) throughout the range $\mZp>1\TeV$ for $\mphi = 50\GeV$.
Below $\mZp=1\TeV$, the efficiency decreases rapidly because of the effect
on the acceptance of the kinematic requirements.
Heavier \mphi values
correspond to a less boosted signature and therefore are selected with a higher efficiency.
For $\mZp > 2$\TeV, the efficiency for the other \mphi samples is approximately 10--15\% (1--5\%)
higher in the electron (muon) channels than for the $\mphi = 50\GeV$ scenario,
where the range of values reflects the variation with \mZp.
For $\mZp < 1\TeV$, the contribution from boosted events is not significant
and the efficiency is similar for all values of $\mphi$ considered.

\section{Background estimation}

Most of the SM backgrounds are suppressed by requiring four isolated high-quality lepton candidates.
As discussed above, the dominant background is from $\cPZ\cPZ$ events decaying into four leptons.
Other backgrounds originate from top quark events with two genuine leptons
and two lepton candidates arising from misidentified jets, and
from $\PW\PW$ ($\PW\cPZ$) events that contain two (one) misidentified or
nonprompt leptons from jets.
In the case of triboson production, there may be four genuine leptons in the event.
These backgrounds are estimated using MC simulation.

The contribution from events with more than two leptons arising from misidentified jets is expected to be small
because this analysis requires four isolated leptons in the final state.
This background is estimated using the ``misidentification rate'' method described in Ref.~\cite{EXO-12-061}.
The misidentification rate measured as a function of electron \et in the barrel and endcap
is applied to events with electron candidates passing the trigger but failing the full selection.
The contribution from jet backgrounds estimated using this procedure is found to be negligible.

\begin{figure*}[htb]
\centering
\includegraphics[width=0.85\textwidth]{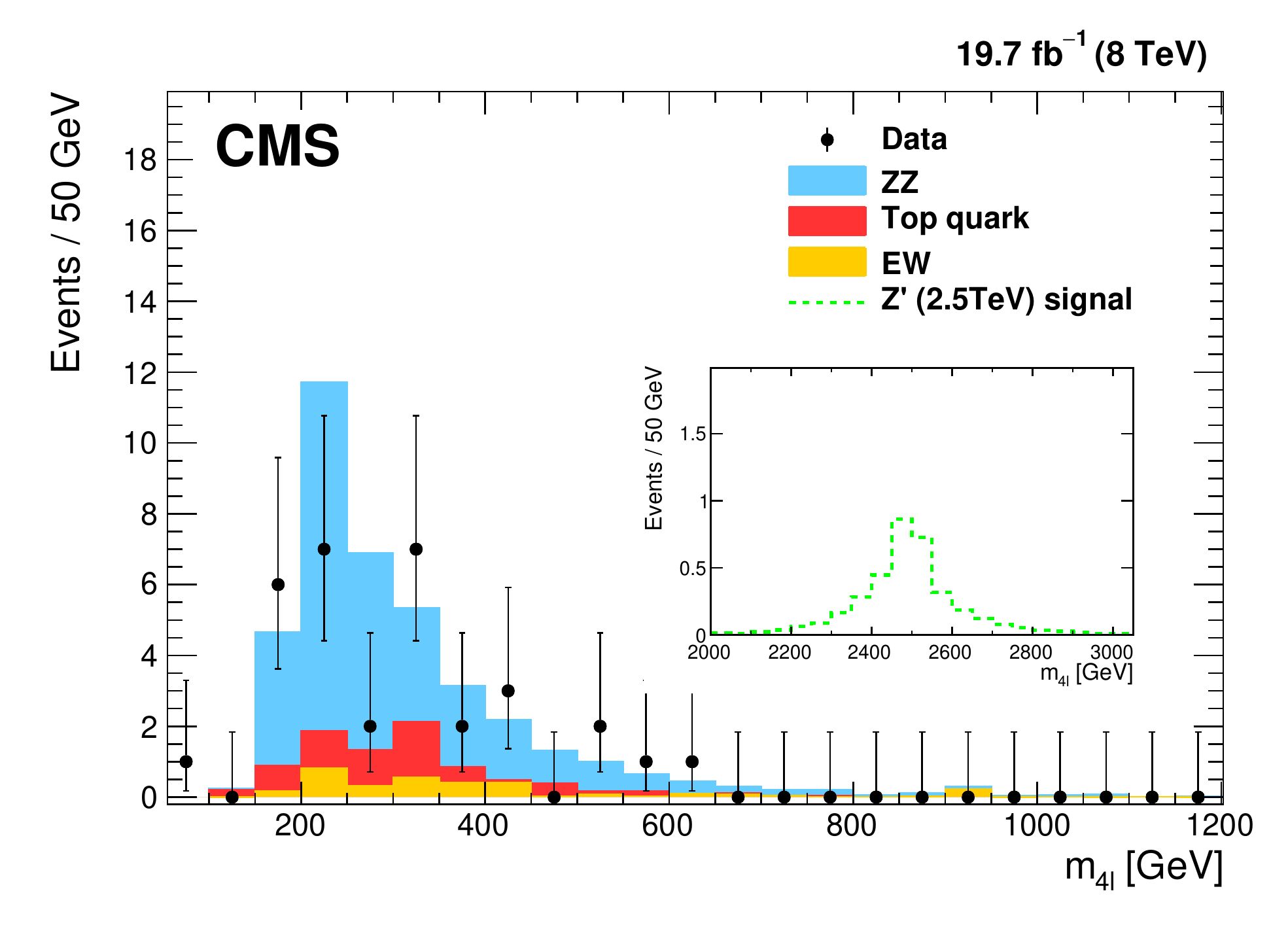}
\caption{
  The \mfl spectrum for the combination of the five studied
  channels. The points with vertical bars represent the data and the associated statistical uncertainties;
  the histograms represent the expectations from SM processes;
  ``Top quark'' denotes the sum of the events for \ttbar, $\PQt\PW$,
 $\ttbar\cPZ$~processes; ``EW'' denotes the sum of the events from
  $\PW\PW, \PW\cPZ, \PW\PW\gamma, \PW\PW\cPZ, \PW\cPZ\cPZ$, and $\cPZ\cPZ\cPZ$ processes. The inset shows the expectation from the benchmark model for a signal at \mZp = 2.5\TeV with $\mphi = 50\GeV$.
}
\label{fig:controlPlots}
\end{figure*}

Figure~\ref{fig:controlPlots} shows the four-lepton invariant mass (\mfl) distribution for selected events.
The number of observed events and estimated backgrounds are summarized in Table~\ref{tab:Nobs}.
As shown in the figure and table,
the distribution of observed events is in agreement with the expected backgrounds.
The table shows two different mass ranges.
In the region $\mfl>1\TeV$, the backgrounds from SM processes
are very small, typically less than one event.

\begin{table*}[htb]
\topcaption{
Summary of the observed yield and expected backgrounds for all channels, where
$N_\text{obs}$ is the number of observed events in data.
The total background ($N_\text{tot}$) is the sum of three different backgrounds
that are estimated using MC simulations;
$N_{\cPZ\cPZ}$ refers to the background from $\cPZ\cPZ$ events;
$N_{\PQt}$ is the background from \ttbar, single top quark,
and $\ttbar\cPZ$~production; $N_\mathrm{EW}$
is the background from $\PW\PW$ and $\PW\cPZ$, and triple gauge boson production.
The quoted uncertainties are statistical only.
}
\centering
{
\begin{tabular}{lccccccc}
\hline
\multirow{3}{*}{Channel}&  \multicolumn{5}{c}{$0.1 < \mfl < 1.0$\TeV} & \multicolumn{2}{c}{$\mfl > 1.0$\TeV} \\
\cline{2-8}
& \multirow{2}{*}{$N_\text{obs}$} & \multicolumn{4}{c}{SM~backgrounds} & \multirow{2}{*}{$N_\text{obs}$} & \multirow{2}{*}{$N_\text{tot}$} \\
& & $N_{\cPZ\cPZ}$ & $N_{\PQt}$ & $N_\mathrm{EW}$ & $N_\text{tot}$ & &  \\
\hline
$\PZpr\to\mumumumu$   & 3  & 4.9 $\pm$ 0.3 & 0.9 $\pm$ 0.5 & ----          &  5.9 $\pm$ 0.6 & 0 & ---- \\
$\PZpr\to\mumumue$    & 6  & 0.4 $\pm$ 0.1 & 1.3 $\pm$ 0.6 & 1.2 $\pm$ 0.3 &  2.9 $\pm$ 0.7 & 0 & ---- \\
$\PZpr\to\mumuee$     & 12 & 9.3 $\pm$ 0.4 & 3.0 $\pm$ 1.5 & 1.2 $\pm$ 0.3 & 13.5 $\pm$ 1.6 & 0 & 0.1 $\pm$ 0.1 \\
$\PZpr\to\mueee$      & 2  & 0.2 $\pm$ 0.1 & 0.4 $\pm$ 0.1 & 0.6 $\pm$ 0.2 &  1.2 $\pm$ 0.2 & 0 & 0.1 $\pm$ 0.1 \\
$\PZpr\to\eeee$       & 9  & 15.0$\pm$ 0.5 & 0.2 $\pm$ 0.1 & 0.2 $\pm$ 0.1 & 15.4 $\pm$ 0.5 & 0 & 0.2 $\pm$ 0.1 \\
\hline
Combined		& 32 & 29.9$\pm$ 0.7 & 5.7 $\pm$ 1.9 & 3.3 $\pm$ 0.5 & 38.9 $\pm$ 2.1 & 0 & 0.4 $\pm$ 0.2 \\
\hline
\end{tabular}
}
\label{tab:Nobs}

\end{table*}

\section{Results}

No excess of events is observed in the data sample compared to the SM expectations and
exclusion limits at 95\% confidence level (CL) are calculated in the
context of the benchmark model.
The signal region consists of events with four leptons ($\Pe$ or $\Pgm$) with $\abs{\eta}<2.4$:
the two leading (subleading) leptons are required to have $\pt>45\,(30)\GeV$.
A Bayesian approach is adopted with a likelihood function defined with a signal strength modifier,
a prior probability, and a set of nuisance parameters.
The prior probability distribution for the signal cross section is positive and uniform,
since this is known to result in good frequentist coverage properties.
The systematic uncertainties associated with the backgrounds, selection efficiency
and luminosity are treated as nuisance parameters with log-normal prior distributions~\cite{PDG}.
A limit on the signal contribution is derived by interpreting the likelihood function
as a probability distribution and integrating over this.
The coverage of the 95\% CL assigned to the limit has been checked using
a Markov chain Monte Carlo method.

The systematic uncertainties are dominated by the uncertainties in the background estimates and in the lepton selection efficiencies.
The uncertainty in the MC estimation of the main background cross section ($\cPZ\cPZ$ and \ttbar)
arising from higher-order QCD corrections and choice of PDFs is 15\%.
In order to be conservative, we choose to double this figure and assign an uncertainty of 30\% from
this source.
The systematic uncertainty in the muon selection efficiency including reconstruction, identification,
and isolation is 0.5\%~\cite{CMS-PAPER-MUO-10-004}.
The uncertainties in the electron selection efficiency are 0.7\% (0.6\%) for
electrons below 100\GeV in EB (EE) and 1.4\% (0.4\%) for electrons above 100\GeV in EB\,(EE)~\cite{EXO-12-061}.
The uncertainties due to the lepton efficiency in both signal and background
yields vary between 2.2\% and 2.7\% as a function of \mfl.
Including the effect of the merged lepton signature, a total uncertainty of 10\% in the
event selection efficiency is assigned for each channel.
The impact of the uncertainty in the electron energy scale on signal (background) yield is 1\% (0.5\%)~\cite{EXO-12-061}.
Uncertainties in the muon momentum scale and mass resolutions are below 0.1\%~\cite{CMS-PAPER-MUO-10-004}.
The uncertainty in the integrated luminosity is assigned to be 2.6\%~\cite{CMS-PAS-LUM-13-001}.
In this analysis, the statistical uncertainties are dominant and the systematic uncertainties
have a small impact on the results.
We tested the robustness of the limits by doubling the values assumed for the systematic uncertainties.
We observed a negligible change in the calculated limits,
and conclude that the limits are insensitive to any underestimation of the systematic uncertainties.

\begin{figure}[htbp]
\centering
\includegraphics[width=0.495\textwidth]{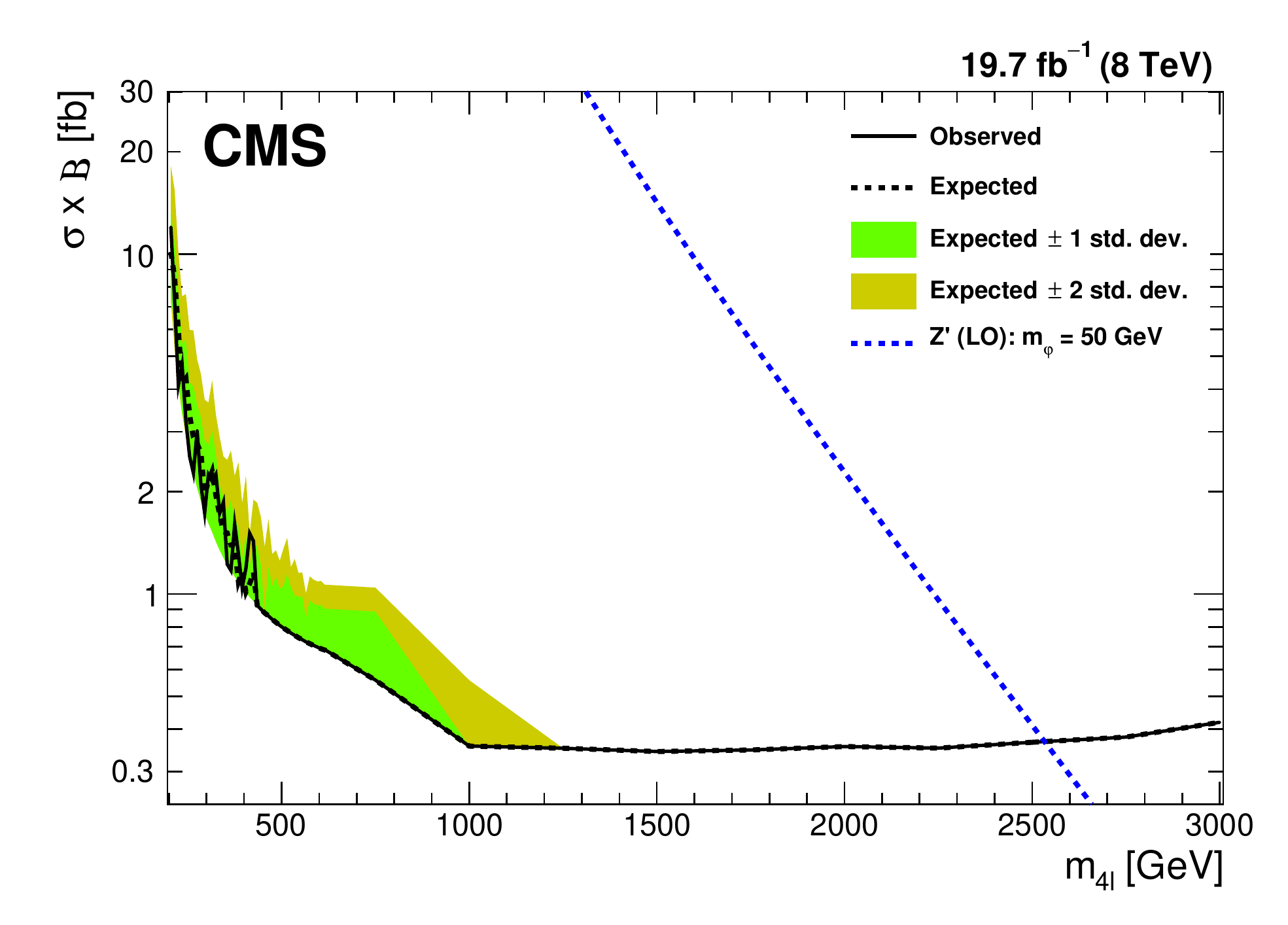}
\includegraphics[width=0.495\textwidth]{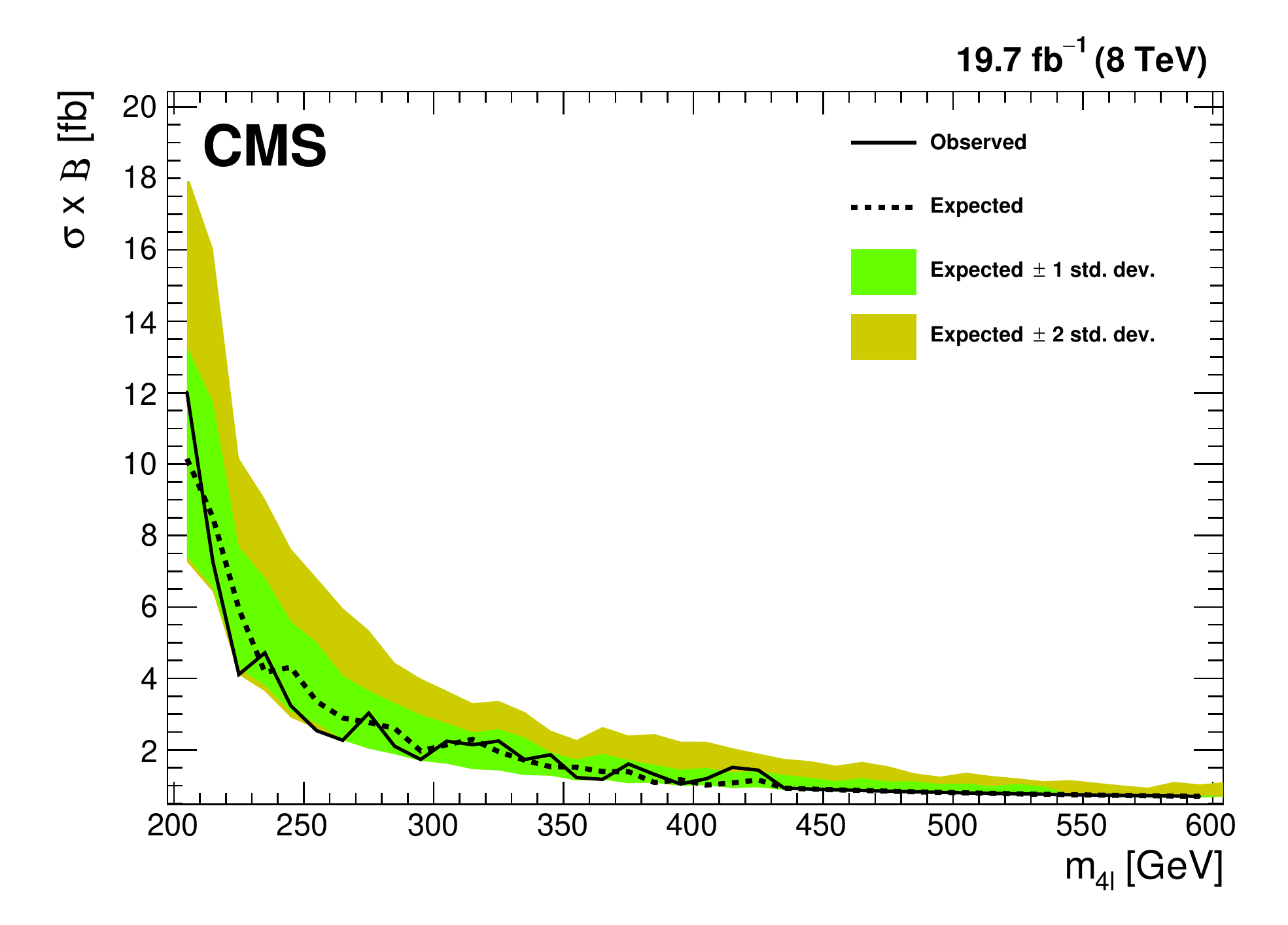}
\caption{
   The 95\% CL upper limit on the cross section times branching fraction
   as a function of \mfl for the combination of the five channels: full mass range (\cmsLeft)
   and expanded view of the low mass region (\cmsRight).
   The shaded green (yellow) band indicates the one (two) sigma uncertainty in the expected limits.
   The blue dashed line represents the theoretical predictions for the benchmark
   model~\cite{HSLee:FourLepton} for $\mphi = 50\GeV$.
}
\label{fig:limit}
\end{figure}

Limits on the product of cross section and branching fraction are set in the context
of the benchmark model as a function of \mfl.
The natural width of the \PZpr resonance is assumed to be smaller than the mass resolution
of the detector in all channels.
The detector resolution in the $\mumumumu$ channel varies from 1.1\% at $\mfl=250\GeV$ to
7.5\% at $\mfl=3\TeV$, and it has a constant value of about 0.6\% over this range in the $\eeee$ channel.
In the limit calculation, we set the mass window to be six times the mass resolution
centred around the signal mass point considered.
A counting experiment is performed for the limit calculation.
Figure~\ref{fig:limit} shows the upper limits on the product of the cross section and branching fraction,
for the combination of all five channels, for the mass range considered in the benchmark model
of Ref.~\cite{HSLee:FourLepton}.
In the framework of this model, we translate these cross section upper limits into lower
limits on the \PZpr boson mass.
For the combination of the five channels, the value obtained for this lower mass limit is 2.5\TeV.
The black solid (dashed) line indicates the observed (expected) 95\% CL upper limits,
the inner (outer) band indicates the $\pm1\,(2)$ standard deviation uncertainty in the expected limits,
and the blue dashed line shows the theoretical \PZpr cross section for $\mphi = 50\GeV$.
This theoretical cross section is calculated under the benchmark model assumption
that the branching fraction $\mathcal{B}(\varphi \to \ell\ell') = 100\%$.
In the region above the 1--1.5\TeV, the bands are not visible since backgrounds are negligible here.

Table~\ref{tab:limit2D} shows the exclusion limit on \mZp for the five separate channels and for the combination.
Results are presented for the benchmark assumption $\mphi = 50\GeV$, and for the five different
values of the ratio \mphi/\mZp assumed for the generated signal samples, taking into account the event
selection efficiencies described above. The predicted cross sections decrease as the ratio \mphi/\mZp increases.
The contribution of the merged lepton signature also decreases, resulting in an overall efficiency increase.
Therefore the scenarios with \mphi/\mZp = 5, 10, 20, 30 and 40\% of \mZp, give slightly higher limits than the $\mphi = 50\GeV$ scenario.

\begin{table*}[htpb]
\topcaption{
 The 95\% CL lower limits (in \TeV) on \mZp for the five separate channels and for their combination.
 Results are presented for the benchmark assumption $\mphi = 50\GeV$, and for the five different values of the ratio \mphi/\mZp.}
\centering
{
\begin{tabular}{lcccccc}
\hline
\mphi  & 50\GeV & $0.05\mZp$ & $0.1\mZp$ & $0.2\mZp$ & $0.3\mZp$ & $0.4\mZp$ \\
\hline
$\mumumumu$	& 1.7 & 1.6 & 1.7 & 1.7 & 1.7 & 1.7  \\
$\mumumue$	& 2.0 & 2.0 & 2.1 & 2.1 & 2.1 & 2.1  \\
$\mumuee$	& 2.4 & 2.4 & 2.5 & 2.5 & 2.5 & 2.5  \\
$\mueee$	& 2.0 & 2.0 & 2.1 & 2.1 & 2.1 & 2.1  \\
$\eeee$		& 1.7 & 1.7 & 1.7 & 1.7 & 1.7 & 1.7  \\
\hline
Combined 	& 2.5 & 2.6 & 2.6 & 2.6 & 2.6 & 2.6  \\
\hline
\end{tabular}
}
\label{tab:limit2D}

\end{table*}

\section{Summary}

Results have been presented from a search for heavy narrow resonances decaying into four-lepton
final states via intermediate scalar particles $\varphi$,
where the branching fraction of $\varphi \to \ell\ell$ ($\ell = \Pe$ or $\Pgm$) is set to 1.
These results are based on a sample of proton-proton collision data at $\sqrt{s} = 8\TeV$,
corresponding to an integrated luminosity of 19.7\fbinv.
The four-lepton invariant mass spectra are consistent with the standard model predictions.
Masses of Z' bosons have been excluded at 95\% confidence level for a specific benchmark model
with $\mphi = 50\GeV$, and for five different assumptions for the ratio \mphi/\mZp (\mphi/\mZp = 5, 10, 20, 30 and 40\%).
Five decay channels ($\mumumumu, \mumumue, \mumuee$, $\mueee, \eeee$) are considered in this analysis.
Combining all channels, a lower limit on the \PZpr mass of $2.5\TeV$ is obtained for the benchmark model,
and $2.6\TeV$ for each of the models assuming a fixed ratio between \mphi and \mZp.
This is the first result to constrain a leptophobic \PZpr resonance in the four-lepton channel.

\begin{acknowledgments}
We congratulate our colleagues in the CERN accelerator departments for the excellent performance of the LHC and thank the technical and administrative staffs at CERN and at other CMS institutes for their contributions to the success of the CMS effort. In addition, we gratefully acknowledge the computing centres and personnel of the Worldwide LHC Computing Grid for delivering so effectively the computing infrastructure essential to our analyses. Finally, we acknowledge the enduring support for the construction and operation of the LHC and the CMS detector provided by the following funding agencies: BMWFW and FWF (Austria); FNRS and FWO (Belgium); CNPq, CAPES, FAPERJ, and FAPESP (Brazil); MES (Bulgaria); CERN; CAS, MoST, and NSFC (China); COLCIENCIAS (Colombia); MSES and CSF (Croatia); RPF (Cyprus); SENESCYT (Ecuador); MoER, ERC IUT, and ERDF (Estonia); Academy of Finland, MEC, and HIP (Finland); CEA and CNRS/IN2P3 (France); BMBF, DFG, and HGF (Germany); GSRT (Greece); OTKA and NIH (Hungary); DAE and DST (India); IPM (Iran); SFI (Ireland); INFN (Italy); MSIP and NRF (Republic of Korea); LAS (Lithuania); MOE and UM (Malaysia); BUAP, CINVESTAV, CONACYT, LNS, SEP, and UASLP-FAI (Mexico); MBIE (New Zealand); PAEC (Pakistan); MSHE and NSC (Poland); FCT (Portugal); JINR (Dubna); MON, RosAtom, RAS, RFBR and RAEP (Russia); MESTD (Serbia); SEIDI and CPAN (Spain); Swiss Funding Agencies (Switzerland); MST (Taipei); ThEPCenter, IPST, STAR, and NSTDA (Thailand); TUBITAK and TAEK (Turkey); NASU and SFFR (Ukraine); STFC (United Kingdom); DOE and NSF (USA).

\hyphenation{Rachada-pisek} Individuals have received support from the Marie-Curie programme and the European Research Council and EPLANET (European Union); the Leventis Foundation; the A. P. Sloan Foundation; the Alexander von Humboldt Foundation; the Belgian Federal Science Policy Office; the Fonds pour la Formation \`a la Recherche dans l'Industrie et dans l'Agriculture (FRIA-Belgium); the Agentschap voor Innovatie door Wetenschap en Technologie (IWT-Belgium); the Ministry of Education, Youth and Sports (MEYS) of the Czech Republic; the Council of Science and Industrial Research, India; the HOMING PLUS programme of the Foundation for Polish Science, cofinanced from European Union, Regional Development Fund, the Mobility Plus programme of the Ministry of Science and Higher Education, the National Science Center (Poland), contracts Harmonia 2014/14/M/ST2/00428, Opus 2014/13/B/ST2/02543, 2014/15/B/ST2/03998, and 2015/19/B/ST2/02861, Sonata-bis 2012/07/E/ST2/01406; the Thalis and Aristeia programmes cofinanced by EU-ESF and the Greek NSRF; the National Priorities Research Program by Qatar National Research Fund; the Programa Clar\'in-COFUND del Principado de Asturias; the Rachadapisek Sompot Fund for Postdoctoral Fellowship, Chulalongkorn University and the Chulalongkorn Academic into Its 2nd Century Project Advancement Project (Thailand); and the Welch Foundation, contract C-1845.
\end{acknowledgments}

\ifthenelse{\boolean{cms@external}}{\vspace*{2ex}}{}
\bibliography{auto_generated}

\cleardoublepage \appendix\section{The CMS Collaboration \label{app:collab}}\begin{sloppypar}\hyphenpenalty=5000\widowpenalty=500\clubpenalty=5000\input{EXO-14-006-authorlist.tex}\end{sloppypar}
\end{document}

%% file: EXO-14-006-authorlist.tex
\textbf{Yerevan Physics Institute,  Yerevan,  Armenia}\\*[0pt]
V.~Khachatryan, A.M.~Sirunyan, A.~Tumasyan
\vskip\cmsinstskip
\textbf{Institut f\"{u}r Hochenergiephysik,  Wien,  Austria}\\*[0pt]
W.~Adam, E.~Asilar, T.~Bergauer, J.~Brandstetter, E.~Brondolin, M.~Dragicevic, J.~Er\"{o}, M.~Flechl, M.~Friedl, R.~Fr\"{u}hwirth\cmsAuthorMark{1}, V.M.~Ghete, C.~Hartl, N.~H\"{o}rmann, J.~Hrubec, M.~Jeitler\cmsAuthorMark{1}, A.~K\"{o}nig, I.~Kr\"{a}tschmer, D.~Liko, T.~Matsushita, I.~Mikulec, D.~Rabady, N.~Rad, B.~Rahbaran, H.~Rohringer, J.~Schieck\cmsAuthorMark{1}, J.~Strauss, W.~Treberer-Treberspurg, W.~Waltenberger, C.-E.~Wulz\cmsAuthorMark{1}
% ...
\vskip\cmsinstskip
% ...
1:~~Also at Vienna University of Technology, Vienna, Austria\\
% ...


%% file: EXO-14-006_temp.bbl
\providecommand{\href}[2]{#2}\begingroup\raggedright\begin{thebibliography}{10}%
\makeatletter
\providecommand{\hrefCMSnoop }[0]{\@secondoftwo}%
\makeatother
\providecommand{\doi}{\texttt{doi:}\begingroup \urlstyle{tt}\Url}

\bibitem{GUT1}
\hrefCMSnoop {}{A.~Leike, ``The phenomenology of extra neutral gauge bosons'',}
  \textit{ Phys. Rept.} \textbf{ 317} (1999) 143,
  \href{http://dx.doi.org/10.1016/S0370-1573(98)00133-1}{\doi{10.1016/S0370-1573(98)00133-1}},
  \href{http://www.arXiv.org/abs/hep-ph/9805494}{\texttt{arXiv:hep-ph/9805494}}.

\bibitem{GUT2}
\hrefCMSnoop {}{J.~L. Hewett and T.~G. Rizzo, ``Low-energy phenomenology of
  superstring-inspired e6 models'',} \textit{ Phys. Rept.} \textbf{ 183} (1989)
  193,
  \href{http://dx.doi.org/10.1016/0370-1573(89)90071-9}{\doi{10.1016/0370-1573(89)90071-9}}.

\bibitem{KK1}
\hrefCMSnoop {}{K.~R. Dienes, E.~Dudas, and T.~Gherghetta, ``Extra spacetime
  dimensions and unification'',} \textit{ Phys. Lett. B} \textbf{ 436} (1998)
  55,
  \href{http://dx.doi.org/10.1016/S0370-2693(98)00977-0}{\doi{10.1016/S0370-2693(98)00977-0}},
  \href{http://www.arXiv.org/abs/hep-ph/9803466}{\texttt{arXiv:hep-ph/9803466}}.

\bibitem{KK2}
\hrefCMSnoop {}{T.~Appelquist, H.-C. Cheng, and B.~A. Dobrescu, ``Bounds on
  universal extra dimensions'',} \textit{ Phys. Rev. D} \textbf{ 64} (2001)
  035002,
  \href{http://dx.doi.org/10.1103/PhysRevD.64.035002}{\doi{10.1103/PhysRevD.64.035002}},
  \href{http://www.arXiv.org/abs/hep-ph/0012100}{\texttt{arXiv:hep-ph/0012100}}.

\bibitem{Ext1}
\hrefCMSnoop {}{L.~Randall and R.~Sundrum, ``An alternative to
  compactification'',} \textit{ Phys. Rev. Lett.} \textbf{ 83} (1999) 4690,
  \href{http://dx.doi.org/10.1103/PhysRevLett.83.4690}{\doi{10.1103/PhysRevLett.83.4690}},
  \href{http://www.arXiv.org/abs/hep-ph/9906064}{\texttt{arXiv:hep-ph/9906064}}.

\bibitem{Ext2}
\hrefCMSnoop {}{L.~Randall and R.~Sundrum, ``A large mass hierarchy from a
  small extra dimension'',} \textit{ Phys. Rev. Lett.} \textbf{ 83} (1999)
  3370,
  \href{http://dx.doi.org/10.1103/PhysRevLett.83.3370}{\doi{10.1103/PhysRevLett.83.3370}},
  \href{http://www.arXiv.org/abs/hep-ph/9905221}{\texttt{arXiv:hep-ph/9905221}}.

\bibitem{ATLASdijet}
\hrefCMSnoop {}{{ATLAS Collaboration}, ``{Search for new phenomena in the dijet
  mass distribution using $pp$ collision data at $\sqrt{s} =$ 8 TeV with the
  ATLAS detector}'',} \textit{ Phys. Rev. D} \textbf{ 91} (2015) 052007,
  \href{http://dx.doi.org/10.1103/PhysRevD.91.052007}{\doi{10.1103/PhysRevD.91.052007}},
  \href{http://www.arXiv.org/abs/1407.1376}{\texttt{arXiv:1407.1376}}.

\bibitem{ATLASdijet13}
\hrefCMSnoop {}{{ATLAS Collaboration}, ``{Search for new phenomena in dijet
  mass and angular distributions from $pp$ collisions at $\sqrt{s} = $ 13 TeV
  with the ATLAS detector}'',} \textit{ Phys. Lett. B} \textbf{ 754} (2016)
  302,
  \href{http://dx.doi.org/10.1016/j.physletb.2016.01.032}{\doi{10.1016/j.physletb.2016.01.032}},
  \href{http://www.arXiv.org/abs/1512.01530}{\texttt{arXiv:1512.01530}}.

\bibitem{CMSdijet}
\hrefCMSnoop {}{{CMS Collaboration}, ``{Search for narrow resonances using the
  dijet mass spectrum in $pp$ collisions at $\sqrt{s} =$ 8 TeV}'',} \textit{
  Phys. Rev. D} \textbf{ 87} (2013) 114015,
  \href{http://dx.doi.org/10.1103/PhysRevD.87.114015}{\doi{10.1103/PhysRevD.87.114015}},
\href{http://www.arXiv.org/abs/1302.4794}{\texttt{arXiv:1302.4794}}.

\bibitem{CMSdijet13}
\hrefCMSnoop {}{{CMS Collaboration}, ``Search for narrow resonances decaying to
  dijets in proton-proton collisions at {$\sqrt{s} =$ 13 TeV}'',} \textit{
  Phys. Rev. Lett.} \textbf{ 116} (2016) 071801,
  \href{http://dx.doi.org/10.1103/PhysRevLett.116.071801}{\doi{10.1103/PhysRevLett.116.071801}},
  \href{http://www.arXiv.org/abs/1512.01224}{\texttt{arXiv:1512.01224}}.

\bibitem{ATLASZp1}
\hrefCMSnoop {}{{ATLAS Collaboration}, ``{Search for high-mass dilepton
  resonances in $pp$ collisions at $\sqrt{s} =$ 8 TeV with the ATLAS
  detector}'',} \textit{ Phys. Rev. D} \textbf{ 90} (2014) 052005,
  \href{http://dx.doi.org/10.1103/PhysRevD.90.052005}{\doi{10.1103/PhysRevD.90.052005}},
  \href{http://www.arXiv.org/abs/1405.4123}{\texttt{arXiv:1405.4123}}.

\bibitem{ATLASZp2}
\hrefCMSnoop {}{{ATLAS Collaboration}, ``{Search for high-mass new phenomena in
  the dilepton finale state using proton-proton collisions at $\sqrt{s} =$ 13
  TeV with the ATLAS detector}'',} \textit{ Phys. Lett. B} \textbf{ 761} (2016)
  372,
  \href{http://dx.doi.org/10.1016/j.physletb.2016.08.055}{\doi{10.1016/j.physletb.2016.08.055}},
  \href{http://www.arXiv.org/abs/1607.03669}{\texttt{arXiv:1607.03669}}.

\bibitem{EXO-12-061}
\hrefCMSnoop {}{{CMS Collaboration}, ``Search for physics beyond the standard
  model in dilepton mass spectra in proton-proton collisions at {$\sqrt{s} =
  8\TeV$}'',} \textit{ JHEP} \textbf{ 04} (2015) 025,
  \href{http://dx.doi.org/10.1007/JHEP04(2015)025}{\doi{10.1007/JHEP04(2015)025}},
\href{http://www.arXiv.org/abs/1412.6302}{\texttt{arXiv:1412.6302}}.

\bibitem{EXO-15-005}
\hrefCMSnoop {}{{CMS Collaboration}, ``Search for narrow resonances in dilepton
  mass spectra in proton-proton collisions at {$\sqrt{s} =$ 13\TeV} and
  combination with {8\TeV} data'',} \textit{ Phys. Lett. B} \textbf{ 768}
  (2017) 57,
  \href{http://dx.doi.org/10.1016/j.physletb.2017.02.010}{\doi{10.1016/j.physletb.2017.02.010}},
  \href{http://www.arXiv.org/abs/1609.05391}{\texttt{arXiv:1609.05391}}.

\bibitem{ATLASdiphoton1}
\hrefCMSnoop {}{{ATLAS Collaboration}, ``{Search for resonances in diphoton
  events at $\sqrt{s} = 13$\TeV with the ATLAS detector}'',} \textit{ JHEP}
  \textbf{ 09} (2016) 001,
  \href{http://dx.doi.org/10.1007/JHEP09(2016)001}{\doi{10.1007/JHEP09(2016)001}},
  \href{http://www.arXiv.org/abs/1606.03833}{\texttt{arXiv:1606.03833}}.

\bibitem{CMSdiphoton1}
\hrefCMSnoop {}{{CMS Collaboration}, ``Search for resonant production of
  high-mass photon pairs in proton-proton collisions at {$\sqrt{s} =$ 8 and
  13\TeV}'',} \textit{ Phys. Rev. Lett.} \textbf{ 117} (2016) 051802,
  \href{http://dx.doi.org/10.1103/PhysRevLett.117.051802}{\doi{10.1103/PhysRevLett.117.051802}},
  \href{http://www.arXiv.org/abs/1606.04093}{\texttt{arXiv:1606.04093}}.

\bibitem{CMSdiphoton2}
\hrefCMSnoop {}{{CMS Collaboration}, ``{Search for high-mass diphoton
  resonances in proton-proton collisions at 13\TeV and combination with 8\TeV
  search}'',} \textit{ Phys. Lett. B} \textbf{ 767} (2017) 147,
  \href{http://dx.doi.org/10.1016/j.physletb.2017.01.027}{\doi{10.1016/j.physletb.2017.01.027}},
  \href{http://www.arXiv.org/abs/1609.02507}{\texttt{arXiv:1609.02507}}.

\bibitem{ATLASdib1}
\hrefCMSnoop {}{{ATLAS Collaboration}, ``Search for heavy resonances decaying
  to a {Z} boson and a photon in pp collisions at $\sqrt{s} = 13$ {TeV} with
  the {ATLAS} detector'',} \textit{ Phys. Lett. B} \textbf{ 764} (2017) 11,
  \href{http://dx.doi.org/10.1016/j.physletb.2016.11.005}{\doi{10.1016/j.physletb.2016.11.005}},
  \href{http://www.arXiv.org/abs/1607.06363}{\texttt{arXiv:1607.06363}}.

\bibitem{ATLASdib2}
\hrefCMSnoop {}{{ATLAS Collaboration}, ``{Search for Higgs boson pair
  production in the $hh \rightarrow b\bar{b}\tau\tau, \gamma\gamma W W^{*},
  \gamma\gamma b\bar{b}, b\bar{b}b\bar{b}$ channels with the ATLAS
  detector}'',} \textit{ Phys. Rev. D} \textbf{ 92} (2015) 092004,
  \href{http://dx.doi.org/10.1103/PhysRevD.92.092004}{\doi{10.1103/PhysRevD.92.092004}},
  \href{http://www.arXiv.org/abs/1509.04670}{\texttt{arXiv:1509.04670}}.

\bibitem{ATLASdib3}
\hrefCMSnoop {}{{ATLAS Collaboration}, ``Search for {WZ} resonances in the
  fully leptonic channel using pp collisions at {$\sqrt{s} = 8$ TeV} with the
  {ATLAS} detector'',} \textit{ Phys. Lett. B} \textbf{ 737} (2014) 223,
  \href{http://dx.doi.org/10.1016/j.physletb.2014.08.039}{\doi{10.1016/j.physletb.2014.08.039}},
  \href{http://www.arXiv.org/abs/1406.4456}{\texttt{arXiv:1406.4456}}.

\bibitem{EXO-16-021}
\hrefCMSnoop {}{{CMS Collaboration}, ``{Search for high-mass Z$\gamma$
  resonances in $\Pep\Pem\gamma$ and $\mu^+\mu^-\gamma$ final states in
  proton-proton collisions at $\sqrt{s} = 8$ and 13 TeV}'',} \textit{ JHEP}
  \textbf{ 01} (2017) 076,
  \href{http://dx.doi.org/10.1007/JHEP01(2017)076}{\doi{10.1007/JHEP01(2017)076}},
  \href{http://www.arXiv.org/abs/1610.02960}{\texttt{arXiv:1610.02960}}.

\bibitem{EXO-16-025}
\hrefCMSnoop {}{{CMS Collaboration}, ``Search for high-mass {Z$\gamma$}
  resonances in proton-proton collisions at {$\sqrt{s} = 8$ and 13\TeV} using
  jet substructure techniques'',} (2016).
  \href{http://www.arXiv.org/abs/1612.09516}{\texttt{arXiv:1612.09516}}.
  Submitted to Phys. Lett. B.

\bibitem{EXO-12-053}
\hrefCMSnoop {}{{CMS Collaboration}, ``Search for heavy resonances decaying to
  two {H}iggs bosons in final states containing four b quarks'',} \textit{ Eur.
  Phys. J. C} \textbf{ 76} (2016) 371,
  \href{http://dx.doi.org/10.1140/epjc/s10052-016-4206-6}{\doi{10.1140/epjc/s10052-016-4206-6}},
  \href{http://www.arXiv.org/abs/1602.08762}{\texttt{arXiv:1602.08762}}.

\bibitem{EXO-14-009}
\hrefCMSnoop {}{{CMS Collaboration}, ``{Search for a massive resonance decaying
  into a Higgs boson and a W or Z boson in hadronic final states in
  proton-proton collisions at $\sqrt{s} = 8$ TeV}'',} \textit{ JHEP} \textbf{
  02} (2016) 145,
  \href{http://dx.doi.org/10.1007/JHEP02(2016)145}{\doi{10.1007/JHEP02(2016)145}},
  \href{http://www.arXiv.org/abs/1506.01443}{\texttt{arXiv:1506.01443}}.

\bibitem{ATLASttbar1}
\hrefCMSnoop {}{{ATLAS Collaboration}, ``Search for resonances decaying into
  top-quark pairs using fully hadronic decays in pp collisions with {ATLAS at
  $\sqrt{s} = 7$\TeV}'',} \textit{ JHEP} \textbf{ 01} (2013) 116,
  \href{http://dx.doi.org/10.1007/JHEP01(2013)116}{\doi{10.1007/JHEP01(2013)116}},
  \href{http://www.arXiv.org/abs/1211.2202}{\texttt{arXiv:1211.2202}}.

\bibitem{ATLASttbar2}
\hrefCMSnoop {}{{ATLAS Collaboration}, ``Search for \ttbar resonances using
  lepton-plus-jets events in proton-proton collisions at {$\sqrt{s} = 8$\TeV}
  with the {ATLAS} detector'',} \textit{ JHEP} \textbf{ 08} (2015) 148,
  \href{http://dx.doi.org/10.1007/JHEP08(2015)148}{\doi{10.1007/JHEP08(2015)148}},
  \href{http://www.arXiv.org/abs/1505.0718}{\texttt{arXiv:1505.0718}}.

\bibitem{CMSttbar1}
\hrefCMSnoop {}{{CMS Collaboration}, ``{Search for anomalous \ttbar production
  in the highly-boosted all-hadronic final state}'',} \textit{ JHEP} \textbf{
  09} (2012) 029,
  \href{http://dx.doi.org/10.1007/JHEP09(2012)029}{\doi{10.1007/JHEP09(2012)029}},
\href{http://www.arXiv.org/abs/1204.2488}{\texttt{arXiv:1204.2488}}.

\bibitem{CMSttbar2}
\hrefCMSnoop {}{{CMS Collaboration}, ``Search for resonant $t\bar{t}$
  production in proton-proton collisions at {$\sqrt{s}=8\TeV$}'',} \textit{
  Phys. Rev. D} \textbf{ 93} (2016) 012001,
  \href{http://dx.doi.org/10.1103/PhysRevD.93.012001}{\doi{10.1103/PhysRevD.93.012001}},
  \href{http://www.arXiv.org/abs/1506.03062}{\texttt{arXiv:1506.03062}}.

\bibitem{HSLee:FourLepton}
\hrefCMSnoop {}{V.~Barger and H.-S. Lee, ``{Four-lepton resonance at the Large
  Hadron Collider}'',} \textit{ Phys. Rev. D} \textbf{ 85} (2012) 055030,
  \href{http://dx.doi.org/10.1103/PhysRevD.85.055030}{\doi{10.1103/PhysRevD.85.055030}},
\href{http://www.arXiv.org/abs/1111.0633}{\texttt{arXiv:1111.0633}}.

\bibitem{LFV1}
\hrefCMSnoop {}{Y.~Kuno and Y.~Okada, ``Muon decay and physics beyond the
  standard model'',} \textit{ Rev. Mod. Phys.} \textbf{ 73} (2001) 151,
  \href{http://dx.doi.org/10.1103/RevModPhys.73.151}{\doi{10.1103/RevModPhys.73.151}},
  \href{http://www.arXiv.org/abs/hep-ph/9909265}{\texttt{arXiv:hep-ph/9909265}}.

\bibitem{LFV2}
\hrefCMSnoop {}{A.~de~Gouvea and P.~Vogel, ``Lepton flavor and number
  conservation and physics beyond the standard model'',} \textit{ Fund. Sym. in
  the Era of the LHC} \textbf{ 71} (2013) 75,
  \href{http://dx.doi.org/10.1016/j.ppnp.2013.03.006}{\doi{10.1016/j.ppnp.2013.03.006}},
  \href{http://www.arXiv.org/abs/1303.4097}{\texttt{arXiv:1303.4097}}.

\bibitem{LFV3}
\hrefCMSnoop {}{D.~K. Ghosh, P.~Roy, and S.~Roy, ``Four lepton flavor violating
  signals at the lhc'',} \textit{ JHEP} \textbf{ 05} (2012) 067,
  \href{http://dx.doi.org/10.1007/JHEP05(2012)067}{\doi{10.1007/JHEP05(2012)067}},
  \href{http://www.arXiv.org/abs/1203.0187}{\texttt{arXiv:1203.0187}}.

\bibitem{CMS-PAPER-MUO-10-004}
\hrefCMSnoop {}{{CMS Collaboration}, ``Performance of {CMS} muon reconstruction
  in pp collision events at {$\sqrt{s} = 7$\TeV}'',} \textit{ JINST} \textbf{
  7} (2012) P10002,
  \href{http://dx.doi.org/10.1088/1748-0221/7/10/P10002}{\doi{10.1088/1748-0221/7/10/P10002}},
  \href{http://www.arXiv.org/abs/1206.4071}{\texttt{arXiv:1206.4071}}.

\bibitem{CMS:EGM-13-001}
\hrefCMSnoop {}{{CMS Collaboration}, ``{Performance of electron reconstruction
  and selection with the CMS detector in proton-proton collisions at $\sqrt{s}
  = 8$\TeV}'',} \textit{ JINST} \textbf{ 10} (2015) P06005,
  \href{http://dx.doi.org/10.1088/1748-0221/10/06/P06005}{\doi{10.1088/1748-0221/10/06/P06005}},
\href{http://www.arXiv.org/abs/1502.02701}{\texttt{arXiv:1502.02701}}.

\bibitem{Chatrchyan:2008zzk}
\hrefCMSnoop {}{{CMS Collaboration}, ``The {CMS} experiment at the {CERN}
  {LHC}'',} \textit{ JINST} \textbf{ 3} (2008) S08004,
\href{http://dx.doi.org/10.1088/1748-0221/3/08/S08004}{\doi{10.1088/1748-0221/3/08/S08004}}.

\bibitem{CalcHep}
\hrefCMSnoop {}{A.~Belyaev, N.~Christensen, and A.~Pukhov, ``{CalcHEP 3.4 for
  collider physics within and beyond the Standard Model}'',} \textit{ Comp.
  Phys. Commun.} \textbf{ 184} (2013) 1729,
  \href{http://dx.doi.org/10.1016/j.cpc.2013.01.014}{\doi{10.1016/j.cpc.2013.01.014}},
  \href{http://www.arXiv.org/abs/1207.6082}{\texttt{arXiv:1207.6082}}.

\bibitem{Sjostrand:2006za}
\hrefCMSnoop {}{T.~Sj{\"o}strand, S.~Mrenna, and P.~Skands, ``{PYTHIA} 6.4
  physics and manual'',} \textit{ JHEP} \textbf{ 05} (2006) 026,
  \href{http://dx.doi.org/10.1088/1126-6708/2006/05/026}{\doi{10.1088/1126-6708/2006/05/026}},
\href{http://www.arXiv.org/abs/hep-ph/0603175}{\texttt{arXiv:hep-ph/0603175}}.

\bibitem{GG2ZZ}
\hrefCMSnoop {}{T.~Binoth, N.~Kauerm, and P.~Mertsch, ``{Gluon-induced QCD
  corrections to $\mathrm{pp} \rightarrow \cPZ\cPZ \rightarrow
  \ell\bar{\ell}\ell^\prime\bar{\ell}^\prime$}'',} (2008).
  \href{http://www.arXiv.org/abs/0807.0024}{\texttt{arXiv:0807.0024}}. (2008).

\bibitem{POWHEG}
\hrefCMSnoop {}{S.~Alioli, P.~Nason, C.~Oleari, and E.~Re, ``{NLO vector-boson
  production matched with shower in POWHEG}'',} \textit{ JHEP} \textbf{ 07}
  (2008) 060,
  \href{http://dx.doi.org/10.1088/1126-6708/2008/07/060}{\doi{10.1088/1126-6708/2008/07/060}},
\href{http://www.arXiv.org/abs/0805.4802}{\texttt{arXiv:0805.4802}}.

\bibitem{Madgraph}
\hrefCMSnoop {}{J.~Alwall {et~al.}, ``Madgraph v5: going beyond'',} \textit{
  JHEP} \textbf{ 06} (2011) 128,
  \href{http://dx.doi.org/10.1007/JHEP06(2011)128}{\doi{10.1007/JHEP06(2011)128}},
  \href{http://www.arXiv.org/abs/1106.0522}{\texttt{arXiv:1106.0522}}.

\bibitem{NNLOttbar}
\hrefCMSnoop {}{M.~Czakon, P.~Fiedler, and A.~Mitov, ``Total top-quark
  pair-production cross section at hadron colliders through
  o($\alpha^4_{S}$)'',} \textit{ Phys. Rev. Lett.} \textbf{ 110} (2013) 252004,
  \href{http://dx.doi.org/10.1103/PhysRevLett.110.252004}{\doi{10.1103/PhysRevLett.110.252004}},
  \href{http://www.arXiv.org/abs/1303.6254}{\texttt{arXiv:1303.6254}}.

\bibitem{NLOZZ}
\hrefCMSnoop {}{J.~M. Campbell, R.~K. Ellis, and C.~Williams, ``{Vector boson
  pair production at the LHC}'',} \textit{ JHEP} \textbf{ 07} (2011) 018,
  \href{http://dx.doi.org/10.1007/JHEP07(2011)018}{\doi{10.1007/JHEP07(2011)018}},
  \href{http://www.arXiv.org/abs/1105.0020}{\texttt{arXiv:1105.0020}}.

\bibitem{CTEQ6L}
J.~Pumplin\hrefCMSnoop {}{ {et~al.}, ``{New generation of parton distributions
  with uncertainties from global QCD analysis}'',} \textit{ JHEP} \textbf{ 07}
  (2002) 012,
  \href{http://dx.doi.org/10.1088/1126-6708/2002/07/012}{\doi{10.1088/1126-6708/2002/07/012}},
  \href{http://www.arXiv.org/abs/hep-ph/0201195}{\texttt{arXiv:hep-ph/0201195}}.

\bibitem{Z2stune1}
\hrefCMSnoop {}{{CMS Collaboration}, ``{Study of the underlying event at
  forward rapidity in pp collisions at $\sqrt{s} = 0.9, 2.76$, and 7\TeV}'',}
  \textit{ JHEP} \textbf{ 04} (2013) 072,
  \href{http://dx.doi.org/10.1007/JHEP04(2013)072}{\doi{10.1007/JHEP04(2013)072}},
  \href{http://www.arXiv.org/abs/1302.2394}{\texttt{arXiv:1302.2394}}.

\bibitem{Z2stune2}
\hrefCMSnoop {}{{CMS Collaboration}, ``Event generator tunes obtained from
  underlying event and multiparton scattering measurements'',} \textit{ Eur.
  Phys. J. C} \textbf{ 76} (2016) 155,
  \href{http://dx.doi.org/10.1140/epjc/s10052-016-3988-x}{\doi{10.1140/epjc/s10052-016-3988-x}},
  \href{http://www.arXiv.org/abs/1512.00815}{\texttt{arXiv:1512.00815}}.

\bibitem{Geant4}
\hrefCMSnoop {}{{GEANT4} Collaboration, ``{GEANT4} -- a simulation toolkit'',}
  \textit{ Nucl. Instrum. Meth. A} \textbf{ 506} (2003) 250,
  \href{http://dx.doi.org/10.1016/S0168-9002(03)01368-8}{\doi{10.1016/S0168-9002(03)01368-8}}.

\bibitem{Geant4da}
\hrefCMSnoop {}{J.~Allison {et~al.}, ``{GEANT4} developments and
  applications'',} \textit{ IEEE Trans. Nucl. Sci.} \textbf{ 53} (2006) 270,
  \href{http://dx.doi.org/10.1109/TNS.2006.869826}{\doi{10.1109/TNS.2006.869826}}.

\bibitem{PDG}
\hrefCMSnoop {}{{Particle Data Group}, K.~A. Olive {et~al.}, ``{Review of
  Particle Physics}'',} \textit{ Chin. Phys. C} \textbf{ 38} (2014) 090001,
\href{http://dx.doi.org/10.1088/1674-1137/38/9/090001}{\doi{10.1088/1674-1137/38/9/090001}}.

\bibitem{CMS-PAS-LUM-13-001}
\href {http://cdsweb.cern.ch/record/1598864}{{CMS Collaboration}, ``Cms
  luminosity based on pixel cluster counting - summer 2013 update'',} CMS
  Physics Analysis Summary CMS-PAS-LUM-13-001, 2013.

\end{thebibliography}\endgroup
